\documentclass[a4paper]{jpconf}
\usepackage{graphicx}

\usepackage{amsmath}
\usepackage{amsfonts}
\usepackage{amssymb}

\begin{document}

\title{Effects of inhomogeneities on the expansion of the Universe: a challenge to dark energy?}

\author{Marie-No\"elle C\'el\'erier}

\address{Laboratoire Univers et Th\'eories (LUTH),
Observatoire de Paris, CNRS, Universit\'e Paris Diderot, 5 place Jules Janssen,
92190 Meudon, France}

\ead{marie-noelle.celerier@obspm.fr}

\begin{abstract}

The current standard model of cosmology, the $\Lambda$CDM model, is based
on the homogeneous FLRW solutions of the Einstein equations to which some
perturbations are added to account for the CMB features and structure formation at
large scales. This model fits rather well the observations provided 95\% of the
energy density budget of the Universe should be of an unknown physical nature, i.e.
dark matter and dark energy. Now, the aim of a cosmological model is not merely to
reproduce the observations, but also to give a physical understanding of the
Universe we live in. Moreover, even if the assumption of homogeneity seems to be
more or less valid at large scales, it appears to be in contradiction with
observations at intermediate scales (between the scale of non linear structure formation and that where structures virialize). This is the reason why, during the
last decade, a community of researchers formed whose aim was to look for the best
way to take into account the influence of the inhomogeneities seen in the Universe
and to construct accurate cosmological models which could possibly get rid of the
dark components. This task, which is still in its infancy, is currently progressing
towards promising results. Two types of methods can be found in the literature:
spatial averaging of scalar quantities and use of exact inhomogeneous solutions of
General Relativity. We will give here a brief report of the second one.

\end{abstract}

\section{Introduction} \label{intro}

The current standard model of cosmology is based on an hypothesis, the Cosmological ``Principle'', stating that the Universe is spatially homogeneous beyond some large scale. However, even if the value of this scale is usually claimed to be of the order of 100 Mpc, this assumption has never been proved and is still under debate. A mathematical simplification describing the Universe by a FLRW model {\it at all scales} is therefore commonly used. Up to 1998, this model was Einstein-de Sitter (EdS).

Then, in 1998, it was found that the Supernova Ia (SN Ia) luminosity is dimmer than predicted by the EdS model \cite{Riess1998}. As a consequence this model was ruled out. However, contrary to what was widely advocated at the time, and still is, one cannot put forward accelerated expansion at this stage. Actually, all current evidence for dark energy is indirect, because the only effect of dark energy is on the expansion history of the Universe.

What is usually done in cosmology is: first assume a cosmological model, then fit it to data measuring observables sensitive to the expansion history through the redshift, thus obtain the values of the cosmological parameters {\it of the assumed model}.

In an FLRW cosmology, the expansion rate $H = \dot{a}/a$ is given by
\begin{equation} \label{eq1}
H^2 (z) = H_0^2 \left\{\Omega_{\Lambda} \exp \left[ \int_0^z \frac{{\rm d}z_1}{1+ z_1} 3 \left[1 + w_{\Lambda}(z_1)\right] \right]+ \Omega_k (1+z)^2 + \Omega_M (1+z)^3 + \Omega_R (1+z)^4\right\}.
\end{equation}
It depends on four constants: the fraction of dark energy, $\Omega_{\Lambda}$, the fraction of curvature, $\Omega_k$, the fraction of matter, including dark matter, $\Omega_M$, the fraction of radiation, $\Omega_R$, and one function, $w_{\Lambda}$, which is the equation of state parameter of the dark energy. Since $H^2 (0) = H_0^2$, $\Sigma_i \Omega_i = 1$, removing one constant. $\Omega_k \sim 0$ from the WMAP experiment measuring the CMB power spectrum and $\Omega_R$ is also small from the CMB ($5 \times 10^{-5}$). The two other constants $\Omega_{\Lambda}$ and  $\Omega_M$, and the function $w_{\Lambda}$ are rather well constrained by their effect on the expansion history, which is manifest through various cosmological observables. The result is the ``concordance'' model with $\Omega_{\Lambda} \sim 0.7$, $\Omega_M \sim 0.3$ (with $\sim 0.25$ from dark matter) and $w = -1$.

Is it a good cosmological model? It seems to work since it reproduces nearly all the cosmological data. But 95\% is ``dark'', i.e. not explained. What about a model whose components are so fundamentally not understood and could the accelerated expansion be a mere mirage?

It has been shown shortly after the discovery of the Sn Ia dimming that inhomogeneities in our local Universe can {\it in principle} mimic an accelerated expansion. A general demonstration for models spherically symmetric around the observer has been provided in \cite{Celerier2000}. Various inhomogeneous models with no cosmological constant reproducing the supernova data have also been proposed in \cite{Dabrowski1998,Pascual1999,Celerier2000,Tomita2000}.

To understand intuitively how such an effect can arise, it is worth turning to the very pedagogical model proposed in \cite{Tomita2000}. This model is composed of two regions. An inner homogeneous region with a given density is matched at some redshift to another homogeneous region with a higher density. Both regions {\it decelerate}, but since the inner void expands faster than the outside region because it contains less matter, the observer inside the void, observing a source in the outer region, experiences {\it an apparent acceleration} of the Universe expansion.

Three methods have been proposed to take inhomogeneities into account: i) linear perturbation theory which is valid when {\it both} the curvature and density contrasts remain small (this is not the case in the non-linear regime of structure formation and where the SNe Ia are observed), ii) averaging of scalar quantities, promising and currently improving iii) exact inhomogeneous solutions which are valid at all scales and are exact perturbations of the Friedmann background which they can reproduce as a limit with any precision. This last one is the method we will consider in the following.

\section{One patch Lema\^itre-Tolman (LT) models with $\Lambda=0$} \label{LT}

\subsection{The LT solution}

The Lema\^itre\cite{Lemaitre1933}-Tolman\cite{Tolman1934} (LT) solution is a spherically symmetric and non-static solution of Einstein's equations, with a dust gravitational source. Its metric in comoving coordinates and synchronous time gauge, in units where $c=1$, is
\begin{equation}\label{eq2}
{\rm d} s^2 = {\rm d} t^2 - \frac {{R'}^2}{1 + 2E(r)}{\rm d} r^2 - R^2(t,r)({\rm d}\vartheta^2 + \sin^2\vartheta \, {\rm d}\varphi^2),
\end{equation}
where the prime denotes derivation with respect to the radial coordinate $r$. A first integral of the field equations gives a dynamical equation for $R$:
\begin{equation}\label{eq3}
\dot{R}^2 = 2E + \frac{2M}{R} + \frac{\Lambda}{3} R^2,
\end{equation}
where the dot denotes derivation with respect to the time coordinate $t$. The mass density in energy units, coming from another field equation, is
\begin{equation}\label{eq4}
8\pi G \rho = \frac {2M'}{R^2R'}
\end{equation}
Integrating Eq.(\ref{eq3}) gives
\begin{equation}\label{eq5}
\int\limits_0^R\frac{{\rm d} \widetilde{R}}{\sqrt{2E + 2M/\widetilde{R} +
\frac{1}{3}\Lambda \widetilde{R}^2}} = t- t_B(r).
\end{equation}
This solution exhibits three integration functions: the energy per unit mass of the particles within the comoving shell of radius $r$, $E(r)$, the gravitational mass of the particles in that shell, $M(r)$, and the ``bang time'', $t_B(r)$ meaning that the singularity (bang or crunch) is not simultaneous for different shells.

To reproduce the cosmological data with no dark energy, the cosmological constant, $\Lambda$, is set to zero. Then Eq.(\ref{eq5}) can be solved explicitly and the solutions are

\begin{itemize}

\item when $E < 0$ (elliptic evolution),
\begin{equation}\label{eqa}
   R(t,r) = \frac{M}{(-2E)}(1 - \cos\eta),
\end{equation}
\begin{equation}\label{eqb}
   \eta - \sin\eta = \frac {(-2E)^{3/2}}{M} (t - t_B(r)),
 \end{equation}
where $\eta(t,r)$ is a parameter

\item  when $E = 0$ (parabolic evolution),
 \begin{equation}   \label{eqc}
   R(t,r) = \left[ \frac{9}{2} M (t - t_B(r))^2\right]^{1/3},
 \end{equation}

\item when $E > 0$ (hyperbolic evolution):
\begin{equation}\label{eqd}
   R(t,r) = \frac{M}{2E} (\cosh\eta - 1),
\end{equation}
\begin{equation}\label{eqe}
\sinh\eta - \eta = \frac {(2E)^{3/2}}{M} (t - t_B(r)).
 \end{equation}

\end{itemize}

Note, however, that a given model can exhibit an $E(r)$ function whose sign changes with $r$ while going smoothly through zero. Even if, for simplification sake, the parametric solutions are commonly used with a fixed sign for $E(r)$, one must be aware of this possible property of the LT models.

Since all the above formulas are covariant under coordinate transformations of the form $\tilde{r} = g(r)$, one of the three integration functions can be fixed at our convenience by the choice of $g$. Therefore, each LT solution is fully determined by two arbitrary functions of $r$ and a coordinate choice.

\subsection{Examples of one patch LT models with $\Lambda = 0$}

This class of solutions has been first used to construct void models in which a local LT under-density centered on the observer and a priori parametrized is matched at some redshift to an EdS model. We discuss below two examples of such models.

\begin{figure}
\begin{center}
\includegraphics[width=9cm,angle=0]{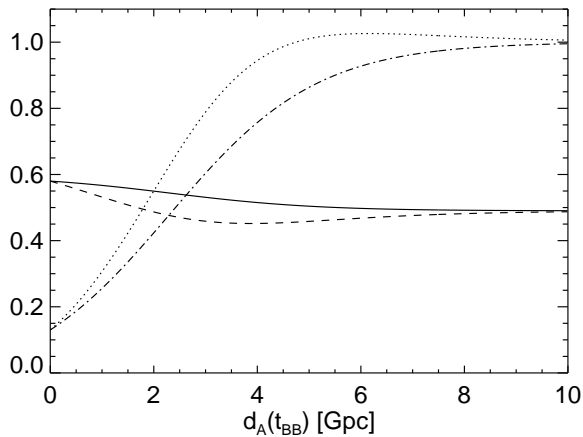}
\end{center}
\caption{\label{fig1} The void for the GBH central observer. Dotted curve = matter density in units of the critical density. Dash-dot curve: $\Omega_M$. Solid curve: $H_T$, the transverse expansion rate. Dashed curve: $H_L$, the longitudinal expansion rate.}
\end{figure}

\begin{figure}
\begin{center}
\includegraphics[width=9cm,angle=0]{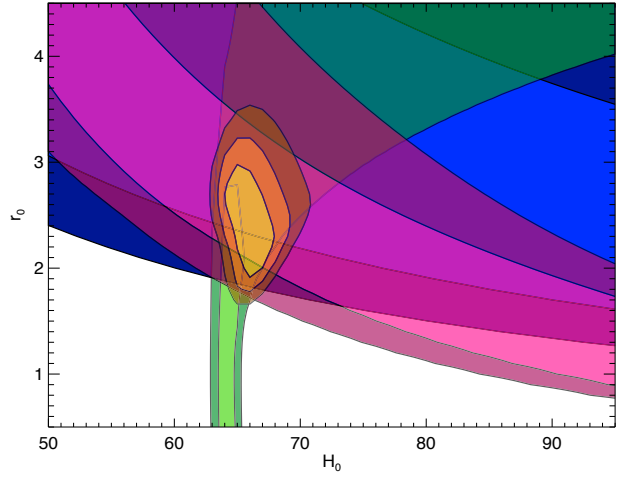}
\end{center}
\caption{\label{fig2}One of the likelihood for the most simple GBH model ($r_0$ characterizes the size of the void): in yellow with 1-, 2-, 3-$\sigma$ contours = the likelihood for the combined data set; in blue, purple and green respectively with 1- and 2-$\sigma$ contours = the likelihood for the individual SN Ia, BAO and CMB data sets. The value of $H_0$ is too low, around 64 (km/s)/Mpc.}
\end{figure}

In \cite{Garcia2008}, the so-called GBH models are specified by their matter content $\Omega_M(r)$ and their expansion rate $H(r)$ and are governed by five free constant parameters. As we said above, they are smoothly matched to an EdS universe at large redshift. These models are fitted to a series of observations (CMB first peak, Large Scale Structures (LSS), Baryon Acoustic Oscillations (BAO), SN Ia data, Hubble Space Telescope measure of $H_0$, age of the globular clusters, gas fraction in clusters) to constrain their parameters. They suggest the possibility that the Universe spherically smoothed around us might be a large (around 2.5 Gpc) void within an EdS universe, with no dark energy. This void is depicted in Fig.\ref{fig1}, with the matter content $\Omega_M$ and the two expansion rates of the LT model, the transverse expansion rate, $H_T \equiv {\dot R}/ R$ and the longitudinal expansion rate $H_L \equiv {\dot R}'/ R'$. However, even if they are able to reproduce most of the observations, they exhibit a value of $H_0$ too low as regards the latest measurements which give $H_0 \sim 74$ (km/s)/Mpc \cite{Riess2011}. This can be seen in Fig.\ref{fig2} where $H_0 \sim 64$ (km/s)/Mpc.

To try to get rid of this drawback a set of five LT void models have been proposed in \cite{Biswas2010}. They have been tested against the full CMB spectrum, BAO, SN and $H_0$.The inclusion of a nonzero overall curvature drastically improves the goodness of fit of these models, bringing them very close to that of the $ \Lambda$CDM model. And, by varying the density profile, the value of $H_0$ has been increased. However, these models have still difficulty exhibiting a value of $H_0$ sufficiently high as regards that of \cite{Riess2011}.

\begin{figure}
\begin{center}
\includegraphics[width=9cm,angle=0]{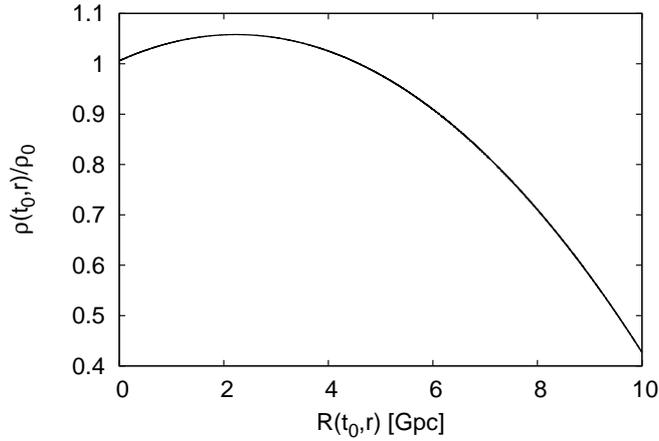}
\end{center}
\caption{\label{fig3} The current density profile of \cite{Celerier2010} does not exhibit a void but a hump. The observer is located in a shallow and wide funnel on top of this hump.}
\end{figure}

In another work \cite{Celerier2010}, an LT model has been fitted to two sets of observables -- the angular diameter distance together with the redshift-space mass density and the angular diameter distance together with the expansion rate -- assumed to have the same form on our past light cone as in the $\Lambda$CDM model. The two left arbitrary LT functions have been determined and have given the mass distribution in space-time. Without any a priori assumption on its form, the current density profile does not exhibit a void but a hump and the observer is located in a shallow and wide funnel on top of this hump, as shown in Fig.\ref{fig3}. Moreover, in this model, the value of $H_0$ is that measured, by construction.

However, single patch LT models with a central observer have been criticized on the ground of being in contradiction with the Copernican Principle. Now, when using such models in cosmology, one should not claim that the observer is at the very centre of any actual spherically symmetric universe. Rather, these models must be considered as mathematical simplifications or preliminary steps. Here, the inhomogeneities are smoothed over angles around us and only their radial component is considered. Note that this is an improvement as regards FLRW models where the inhomogeneities are smoothed over the whole space. Since the exact central location of the observer is mandatory by construction, these models cannot take anisotropies into account. Effects merely designed to test them, such as CMB dipole and low multipoles, kSZ effect, CMB spectral distortion, cosmic parallax etc... can in no way be used to test these models \cite{Celerier2011}.

\section{Swiss-cheese and meat-ball models} \label{Swiss}

\subsection{Example of an LT Swiss-cheese model}

For a more physical representation of our observed Universe, Swiss-cheese models with no dark energy have also been proposed. In \cite{Marra2007} the model is a lattice of LT bubbles with radius $\geq$350 Mpc in an EdS background. Initially, the void at the center of each hole is dominated by negative curvature and a compensating over-density matches smoothly the density and curvature EdS values at the border of the hole. Since the voids expand faster than the cheese, the over-dense regions contract and become thin shells at the borders of the bubbles while under-dense regions turn into emptier voids, eventually occupying most of the volume. An extra very important result is that {\it the evolution} of the voids, bends the photon paths and affects more the photon physics than the geometry of the inhomogeneities.

\subsection{Quasi-spherical Szekeres Swiss-cheese models}

Another exact solution of Einstein's equation has been used to built Swiss-cheese models, the quasi-spherical Szekeres (QSS) solution.

The Szekeres solutions \cite{Szekeres1975} are dust solutions of Einstein's equations with no symmetry at all, i.e. no Killing vector. Their metric in comoving coordinates and synchronous time gauge is
\begin{equation}\label{eq6}
{\rm d} s^2 =  {\rm d} t^2 - {\rm e}^{2\alpha}{\rm d} r^2 - {\rm e}^{2\beta}({\rm d} x^2 + {\rm d} y^2),
\end{equation}
where $\alpha$ and $\beta$ are functions of ($t,r,x,y$) to be determined from the Einstein equations. The $\beta' = 0$ subfamily has not been considered so far in cosmology, since it does not contain the LT solution as a spherically symmetric limit.

The $\beta' \neq 0$ subfamily can be parametrized so that a set of 5 independent functions of $r$ determines each solution. The sign of one of them fixes the geometry of the $\{t = const.\}$ 3-surfaces and the type of evolution (elliptic, parabolic, hyperbolic). The sign of another function determines the geometry of the $\{t=const, r=const\}$ 2-surfaces which can be quasi-spherical, quasi-plane or quasi-hyperbolic. Only the quasi-spherical solution has been found useful in cosmology, since it is the only one which exhibits the three Friedmann limits (hyperbolic, flat and spherical). This solution possesses an origin, not a centre.

Its metric in comoving coordinates and synchronous time gauge can be written, with the parametrization proposed in \cite{Hellaby1996},
\begin{equation}\label{eq7}
{\rm d} s^2 =  {\rm d} t^2 - \frac{(\Phi' - \Phi {  E}'/ {  E})^2}
{1 - k} {\rm d} r^2 - \frac{\Phi^2}{E^2}({\rm d} x^2 + {\rm d} y^2),
\end{equation}
where $\Phi$ is a function of $t$ and $r$, $k$ is a function of $r$ and $E$ is a function of $r$, $x$ and $y$:
\begin{equation}\label{eq8}
{ E} = \frac{S}{2} \left[ \left( \frac{x-P}{S} \right)^2
+ \left( \frac{y-Q}{S} \right)^2 + 1 \right],
\end{equation}
with $S(r)$, $P(r)$, $Q(r)$, functions of $r$.

Applying the Einstein equations to the metric (\ref{eq7}) and assuming that the energy momentum tensor is that of dust, these equations reduce to the following two:
\begin{equation}\label{eq9}
\dot{\Phi}^2 = \frac{2M}{\Phi} - k + \frac{1}{3} \Lambda
\Phi^2,
\end{equation}
\begin{equation}\label{eq10}
\kappa \rho = 
 \frac{2M' - 6 M {  E}'/{  E}}{\Phi^2 ( \Phi' - \Phi {  E}'/{  E})},
\end{equation}
where $\kappa=8\pi G$, $\Lambda$ is the cosmological constant and $M$ is an arbitrary function of $r$.

As in the LT solution, the bang time function, $t_B(r)$, follows from Eq.(\ref{eq9}):
\begin{equation}\label{eq11}
\int\limits_0^{\Phi}\frac{{\rm d} \widetilde{\Phi}}{\sqrt{- k + 2M /
\widetilde{\Phi} + \frac 1 3 \Lambda \widetilde{\Phi}^2}} =  t - t_B(r).
\end{equation}

Note that Eq.(\ref{eq11}) exhibits the same form as Eq.(\ref{eq5}) of the LT solution. Therefore, for $\Lambda =0$, the parametric solutions are the same, with the Szekeres function $k(r)$ corresponding to the LT function $-2E(r)$, i.e.

\begin{itemize}

\item when $k > 0$ (elliptic evolution),
\begin{equation}\label{eqa}
   \Phi(t,r) = \frac{M}{k}(1 - \cos\eta),
\end{equation}
\begin{equation}\label{eqb}
   \eta - \sin\eta = \frac {k^{3/2}}{M} (t - t_B(r)),
 \end{equation}

\item  when $k = 0$ (parabolic evolution),
 \begin{equation}   \label{eqc}
   \Phi(t,r) = \left[ \frac{9}{2} M (t - t_B(r))^2\right]^{1/3},
 \end{equation}

\item when $k < 0$ (hyperbolic evolution):
\begin{equation}\label{eqd}
   \Phi(t,r) = \frac{M}{k} (1 -\cosh\eta),
\end{equation}
\begin{equation}\label{eqe}
\sinh\eta - \eta = \frac {(-k)^{3/2}}{M} (t - t_B(r)).
 \end{equation}

\end{itemize}

Remember that, as for the LT case, the sign of $k(r)$ can change in a given model.

Since all the formulas given so far are covariant under coordinate transformations of the form $\tilde{r} = g(r)$, this means that one of the functions $k(r)$, $S(r)$, $P(r)$, $Q(r)$, $M(r)$ or $t_B(r)$ can be fixed at our convenience by the choice of $g$. Hence, each QSS solution is fully determined by only five functions of $r$ and a coordinate choice.

To begin with and for simplicity sake, axially symmetric classes of QSS models with $\Lambda=0$ have been used to construct Swiss-cheese models able to reproduce the supernova data \cite{Bolejko2010}. Here, the holes are no more spheres, as in LT Swiss-cheeses, but axially symmetric patches and the cheese is either an EdS or an open FLRW background. The luminosity distance is computed on axially directed null geodesics. Compared to those obtained with corresponding LT Swiss-cheeses, the results of these models are quantitatively different but qualitatively comparable. To reproduce the observational data, structures of at least 500 Mpc are necessary.

\subsection{Meat ball models}

The meat ball models are the opposite of the Swiss-cheeses. The holes are replaced by balls and the cheese by voids which occupy more volume than the over-densities \cite{Kainulainen2009}. Most of the photons travel through the voids, missing these localized over-densities. The most interesting finding is that the lensing probability distribution function is skewed with a mode at demagnified values. It is thus possible that these lensing effects, possibly increased by a selection effect, and also because they are stronger in an homogeneous background without cosmological constant than in a $\Lambda$CDM model, might explain part of the ``dark energy''. But these lensing effects alone cannot explain the full dark energy. However, added to other effects, they can boost LT models or make smaller the volume of voids needed in Swiss-cheeses.

\section{Conclusion and prospects} \label{conclusion}

The first very simple inhomogeneous models used to deal with the dark energy problem and proposed very soon after the discovery of the dimming of the luminosity of the supernovae have been with time replaced by more and more sophisticated ones. These are able to reproduce to a better accuracy not only the supernova observations but nearly all the available cosmological data.

To try to improve the reproduction of our observed local Universe, from FLRW models to axially symmetric Szekeres Swiss-cheeses, the number of symmetries has decreased, aiming at the removal of any of them.

It has been claimed that the geometric, dynamical, lensing effects of inhomogeneities act in the same direction \cite{Kainulainen2009}. Their addition can therefore modify markedly our view of the $\Lambda$CDM universe.

Even if at the end of the day the effects of inhomogeneities should prove insufficient to explain out dark energy, the current results obtained with exact models show that their impact on the value of the cosmological parameters will have to be considered as non-negligible.

Now, to be able to take properly into account all the data available to constrain the models, and in particular LSS, BAO and CMB, perturbation theories applicable to the LT and Szekeres models will have to be developed.

The final aim is to construct the metric of the cosmos directly from observations. Preliminary codes have been developed for LT models \cite{Lu2007,McClure2008,Bolejko2011}, but two main problems remain: the choice of the scale and to get rid of spherical symmetry. This implies a jump to full numerical Relativity which might be the future aim of cosmological investigations.

\end{document}